\documentclass[aps,twocolumn,superscriptaddress,nofootinbib]{revtex4-2}
\usepackage[utf8]{inputenc}
\usepackage[T1]{fontenc}
\usepackage{amsmath,amssymb}
\usepackage{amsthm}
\usepackage[pdftex]{graphicx}
\usepackage{color}
\usepackage{xcolor}
\usepackage{epstopdf}
\usepackage{dcolumn}
\usepackage[pdftex]{hyperref}
\hypersetup{colorlinks, citecolor=blue, linkcolor=blue, urlcolor=blue}
\usepackage{lipsum}
\usepackage[labelformat=simple]{subcaption}
\usepackage{cleveref}
\usepackage{mathtools}
\usepackage{physics}
\usepackage{comment}
\usepackage[export]{adjustbox}
\usepackage{float}
\usepackage[format=plain, justification=raggedright]{caption}

\begin{document}

\title{Violation of the Leggett-Garg inequality for dynamics of a Bose-Einstein condensate in a double-well potential}

\author{Tsubasa Sakamoto}
\affiliation{Department of Physics, Chuo University, 1-13-27 Kasuga, Bunkyo-ku, Tokyo 112-8551, Japan}
\author{Ryosuke Yoshii}
\affiliation{Center for Liberal Arts and Sciences, Sanyo-Onoda City University, 1-1-1 Daigaku-Dori, Sanyo-Onoda, Yamaguchi 756-0884, Japan}
\affiliation{International Institute for Sustainability with Knotted Chiral Meta Matter (WPI-SKCM2), Hiroshima University, Higashi-Hiroshima, Hiroshima 739-8526, Japan}
\author{Shunji Tsuchiya}
\affiliation{Department of Physics, Chuo University, 1-13-27 Kasuga, Bunkyo-ku, Tokyo 112-8551, Japan}

\begin{abstract}
The Leggett-Garg inequality (LGI) serves as a criterion to determine the adherence of macroscopic system dynamics to macrorealism, as introduced by Leggett and Garg. A violation of this inequality implies either the absence of a realistic description of the system or the impossibility of noninvasive measurement. 
In this Letter, we investigate the violation of the LGI for the system of bosons in a double-well potential.
Specifically, we explore the violation of the LGI in the dynamics of bosons in a double-well potential in the Bose-Einstein-condensation (BEC) regime, where the system can be considered as two weakly coupled Bose condensates, and in the single-particle regime to establish the conditions under which the violation of the LGI occurs.
Our analysis reveals that the LGI is violated due to Josephson oscillations, while it remains unviolated in the strong coupling regime, attributed to the self-trapping phenomena. 
Notably, we observe that the violation of the LGI becomes increasingly significant as the particle number increases.
These findings provide valuable insights into the macrorealistic behavior of Bose condensates and highlight the effect of measurements on the dynamics of a macroscopic system.
\end{abstract}

\maketitle

\textit{Introduction}.
How can classical behaviors of objects, 
which obey our intuition about how the macroscopic world behaves, 
be distinguished from non-classical behaviors such as what are described by quantum theory?
The Leggett-Garg inequality (LGI) was proposed to answer this question~\cite{LG,Leggett,FN}. It serves as a test of a concept called {\it macrorealism}, 
which consists of two conditions: macrorealism {\it per se} (MRPS) and noninvasive measurability (NIM). 
MRPS assumes that a macroscopic system with two or more macroscopically distinct states available to it is always in one or the other of these states, 
and NIM assumes that it is possible to determine the state of the system with arbitrary small perturbations on its subsequent dynamics.
A violation of the LGI implies a departure from macrorealism. 
The exploration of the LGI was inspired by the question of whether macroscopic coherence, as illustrated by Schr\"odinger's cat gedanken experiment, can be realized in a laboratory~\cite{cat}.

The violation of the LGI has been studied in several systems, experimentally and theoretically~\cite{FN, Emary, Barbieri, Avis, Kofler, Budroni_2013, Budroni, Moreira, Lambert, Formaggio, Martin, Laura, Matsumura, Mawby}. Recently, experimental LG tests have been achieved in a superconducting flux qubit with genuine macroscopicity, refuting its classical realistic description~\cite{flux}.

The system of Bose-Einstein condensates (BECs) in a double-well potential is suitable for testing the LGI to clarify the boundary between classical and quantum regimes.
The dynamics of a BEC in a double-well potential has been observed in several experiments \cite{Albiez, Levy, Gati, Berrada, Ryu, Shin, Saba, Mukhopadhyay, Vretenar}.
In particular, in the experiment of the Heidelberg group~\cite{Albiez}, Josephson oscillations and macroscopic quantum self-trapping were successfully observed, and the experimental result exhibits agreement with the prediction by the 
Gross-Pitaevskii (GP) equation \cite{Smerzi, Gross, Pitaevskii}. It implies that a population imbalance and a phase difference between the two wells have definite values 
independent of measurements in the presence of sufficiently large number of particles in the double-well potential~\cite{GP}.

In this Letter, we discuss the violation of the LGI for bosons in a double-well potential. 
We find the violation of the LGI in the BEC regime, in which the system can be considered as two weakly coupled BECs, despite the fact that their dynamics are well described by coupled semiclassical equations for definite phase and population differences, which are presumed to be unaffected by measurements.
We reveal that the violation of the LGI is associated with the collapse of the state due to projective measurements in the intermediate time, which necessarily affects the subsequent dynamics.
We find that the critical value of interaction strength for the onset of the violation of the LGI coincides with the one for self-trapping. Furthermore, we observe that the violation of the LGI becomes increasingly significant as the particle number increases.
In the single-particle regime, in which bosons do not interact with each other, the bosons tunneling backwards induced by the measurement are found to be crucial for the violation of the LGI.
\begin{figure*}[t!]
\centering
\begin{subfigure}{0.99\textwidth}
\includegraphics[width=\textwidth]{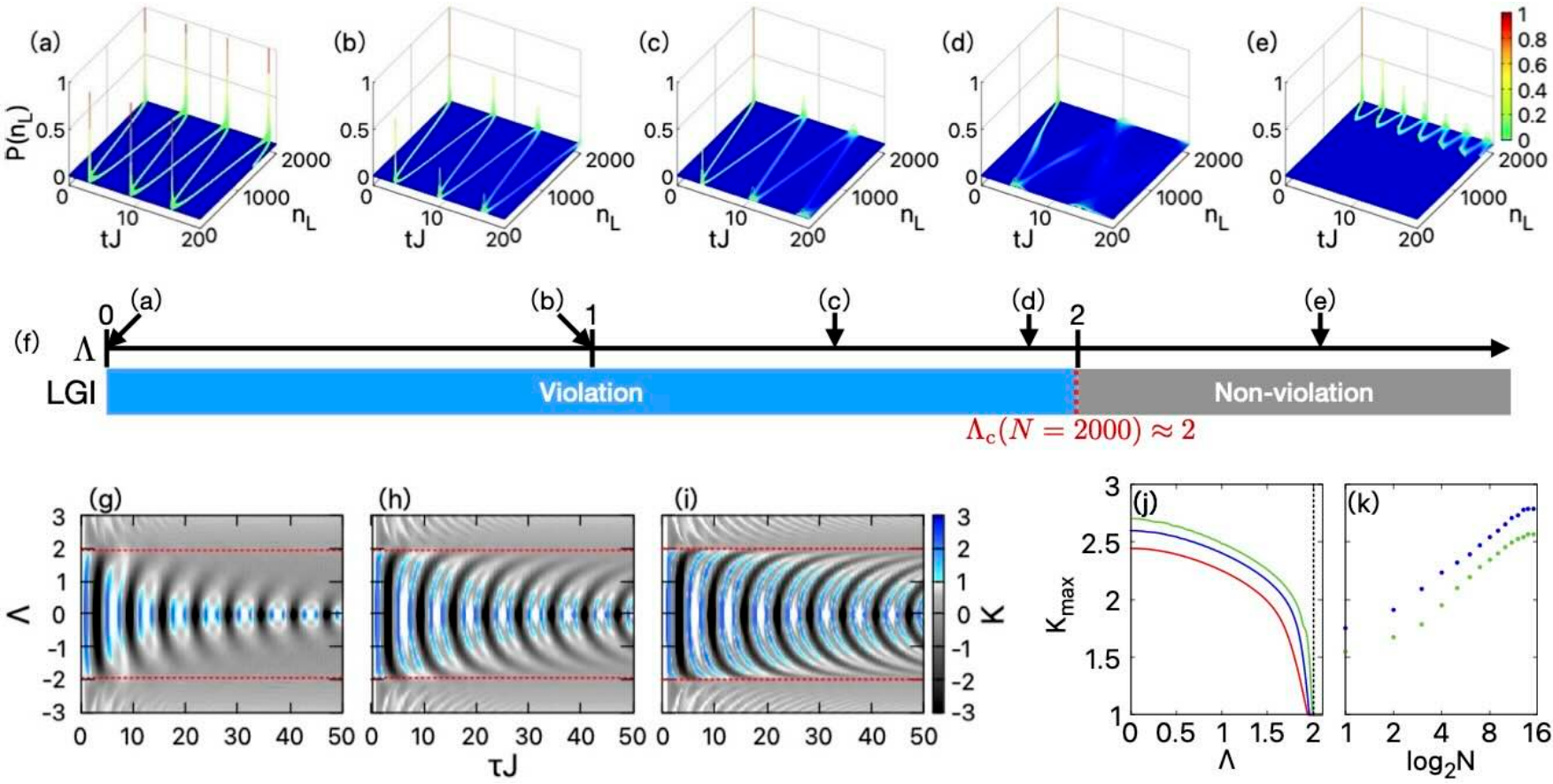}
\phantomsubcaption
\label{fig:te_Lambda0}
\phantomsubcaption
\label{fig:te_Lambda1}
\phantomsubcaption
\label{fig:te_Lambda1.5}
\phantomsubcaption
\label{fig:te_Lambda1.9}
\phantomsubcaption
\label{fig:te_Lambda2.5}
\phantomsubcaption
\label{fig:violation_region}
\phantomsubcaption
\label{fig:LG_Lambda_N100}
\phantomsubcaption
\label{fig:LG_Lambda_N500}
\phantomsubcaption
\label{fig:LG_Lambda_N2000}
\phantomsubcaption 
\label{fig:LGmax_U}
\phantomsubcaption 
\label{fig:LGmax_N}
\end{subfigure}
\caption{(a)\textendash(e) Time evolutions of the occupation probabilities of the left well $P(n_{\rm L})$ for $\Lambda=0$ (a), $1$ (b), $1.5$ (c), $1.9$ (d), and $2.5$ (e). Here, we set $N=2000$.
(f) The light-blue/gray region denotes the violation/nonviolation region. The red dashed line denotes the critical value of the LG violation for $N=2000$.
(g)\textendash(i) Color maps of $K$ for $N=100$ (g), $500$ (h), and $2000$ (i). 
The two red dashed lines in each panel denote the critical values $\pm\Lambda_{\rm c}(N)$. 
(j), (k) Maximal violations of the LGI as functions of $\Lambda$ and $N$, respectively. In (j), the red, blue, and green lines correspond to $N=100$, $500$, and $2000$, respectively. The vertical black dashed line in (j) indicates $\Lambda_{\rm st}^{\rm GP}$. In (k), the blue and green dots correspond to $\Lambda=0$ and $1$, respectively.
}
\label{fig:te_dephasing}
\end{figure*}

\textit{Model}.
The system of bosons in a double-well potential is well described by the Bose-Josephson junction (BJJ) model \cite{Smerzi,Zapata,Raghavan, Ferrini, Hipolito, Xu, Vardi, Zibold, Pezze, Milburn, Sinha,Vardi2}.
Using the Schwinger boson representation $\hat S_x=(\hat a_{\rm L}^\dagger\hat a_{\rm R}+\hat a_{\rm R}^\dagger\hat a_{\rm L})/2$ and $\hat S_z=(\hat n_{\rm L}-\hat n_{\rm R})/2$, 
where $\hat a_{\alpha}$ ($\hat a_{\alpha}^\dagger$) and $\hat n_{\alpha}=\hat a^\dagger_{\alpha}\hat a_{\alpha}$ are annihilation (creation) and number operators of a boson in the well $\alpha={\rm L},{\rm R}$, 
the Hamiltonian of the BJJ model can be written as~\cite{Zibold, Pezze, Milburn}
\begin{align}
\label{H_BJJ}
\hat H_{\rm BJJ}=-J\hat S_x+\frac UN\hat S_z^2,
\end{align}
where $J$ is the hopping strength, $U$ is the on-site interaction, and $N$ is the total number of particles. 
Hereafter, we set $\Lambda=U/J$ and $\hbar=1$.

\textit{Method}.\textemdash
The LGI is formulated for the correlation functions $C_{ij}=\langle Q_iQ_j\rangle$ of dichotomic variables $Q_i=\pm1$, 
which is given as the result of the measurement at time $t_i$, where $i,j=1,2,3$. 
MRPS ensures the existence of the value of the observable $Q_i$ at $t_i$, regardless of whether or not the measurement is performed. 
Thus we can define the joint probabilities $P_{ij}(Q_1,Q_2,Q_3)$, where measurements are performed at $t_i$ and $t_j$, whereas the system is unmeasured at time $t_k$ ($\neq t_i, t_j$). 
Under only MRPS, the joint probability may depend on when measurements are made, 
since measurements at different times can affect the time-evolution differently. 
By adding the NIM condition, however, the subscripts $i$ and $j$ can be dropped since any measurement does not affect the system under NIM.
Then the joint probability $P(Q_1,Q_2,Q_3)$ yields all correlation functions and we obtain the LGI~\cite{LG, FN}.
\begin{equation}
\label{LG}
K\equiv C_{12}+C_{23}-C_{13}\leq1.
\end{equation}

We follow the dynamics of the system numerically by exact diagonalization of the Hamiltonian (\ref{H_BJJ}).  
In evaluation of the LGI, in this work, we define the dichotomic observable as $\hat Q={\rm sgn}(\hat S_z)$, where ${\rm sgn}(x)=1$ if $x\geq0$, otherwise ${\rm sgn}(x)=-1$. 
For the sake of simplicity, we set an initial state at the time $t_1$. 
Throughout this Letter, we focus on the identical time interval and define $\tau=t_2-t_1=t_3-t_2$. 
Then all three two-time correlation functions depend only on the time interval $\tau$. 
Based on the above setting, we calculate $K$ in Eq.~(\ref{LG}) [see also the Supplemental Material (SM) for details on the calculations of the correlation functions].

\textit{BEC regime}.
First, we examine violation of the LGI in the BEC regime ($\Lambda\neq0$ and $N\gg1$), in which the system can be considered as two weakly coupled BECs. Such BECs have been studied by a semiclassical approach based on two coupled GP equations, which predict Josephson oscillation or self-trapping depending on the value of $\Lambda$ and population imbalance at initial time~\cite{Albiez, Levy, Smerzi, Milburn,Julia, Ananikian}. 

\textit{Population dynamics in the BEC regime}. Figures~\ref{fig:te_Lambda0}\textendash\ref{fig:te_Lambda2.5} present the time-evolutions of the occupation probability of the left well $P(n_{\rm L})$ in the case of $N=2000$ with the initial state $|N\rangle_{\rm L}|0\rangle_{\rm R}$.
The peak of the occupation probability oscillates between $n_{\rm L}=0$ and $n_{\rm L}=N$ due to the single-particle tunneling effect when $\Lambda=0$ [Fig.~\ref{fig:te_Lambda0}] A similar coherent oscillation occurs in the presence of an on-site interaction, but it is rather induced by the Josephson effect when $0<\Lambda<\Lambda_{\rm st}$ [Figs.~\ref{fig:te_Lambda1}\textendash\ref{fig:te_Lambda1.9}]. For a strong interaction strength, the Josephson oscillation ceases and self-trapping occurs when $\Lambda>\Lambda_{\rm st}$, where the peak of the occupation probability oscillates keeping the population imbalance [Fig.~\ref{fig:te_Lambda2.5}].
The critical value of interaction strength for the onset of self-trapping $\Lambda_{\rm st}\simeq 2$ agrees well with the value derived in the semiclassical analysis based on the GP equation $\Lambda_{\rm st}^{\rm GP}=2$~\cite{Smerzi}.
\par
If the dynamics obeys coupled GP equations, oscillation of the population imbalance continues without damping~\cite{Smerzi, Raghavan, Ananikian}.
The damping of the oscillation that occurs in the presence of an on-site interaction in Figs.~\ref{fig:te_Lambda1}\textendash\ref{fig:te_Lambda1.9} can be understood as being due to the intrinsic quantum fluctuation in the initial state~\cite{Milburn}. 
That is, since the initial population imbalance is fixed, the initial phase difference between the BECs should be fluctuating due to the uncertainty relation between the population imbalance and phase difference~\cite{Milburn}.
As $n_{\rm L}$ and $n_{\rm R}$ becomes larger, the quantum fluctuation ($\simeq1/\sqrt{n_{{\rm L},{\rm R}}}$~\cite{Gajda}) becomes more negligible~\cite{Pezze, Smerzi}. This is the reason why, with the increase of $N$, the coherent Josephson oscillation persists for longer (see Fig.~S3 in the SM). 
\par
BECs in a double-well potential are considered to possess macroscopic observables for sufficiently large $N$. 
In fact, the damping is suppressed for larger $N$, and the population dynamics is consistent with the semiclassical description (see SM). 
The experimental data in Ref.~\cite{Albiez} have also been shown to be consistent with the prediction of the semiclassical analysis.
Therefore, both the theoretical and experimental studies support the existence of macroscopic wave functions of two BECs with a definite phase difference and population imbalance for sufficiently large $N$.
Upon such considerations, it seems reasonable to expect nonviolation of the LGI.
\par
\textit{Violation of the LGI in the BEC regime}.
Figure~\ref{fig:LG_Lambda_N2000} shows the result of the LG test in the BEC regime for $N=2000$: $K$ is plotted as a function of $\tau$ and $\Lambda$. 
$K$ exhibits periodic oscillations as a function of $\tau$ reflecting the oscillation of the occupation probabilities in Figs.~\ref{fig:te_Lambda0}\textendash\ref{fig:te_Lambda2.5}.
Contrary to the expectation, the LGI is violated near the peaks of the oscillations for the interaction $-\Lambda_{\rm c}<\Lambda< \Lambda_{\rm c}$.
For $|\Lambda|\geq\Lambda_{\rm c}$ the LGI is nonviolated. Remarkably, the critical value $\Lambda_{\rm c}$ that separates violation and nonviolation of the LGI coincides with $\Lambda_{\rm st}$ as illustrated in Fig.~\ref{fig:violation_region}. Thus, the LGI is violated due to Josephson oscillations when $\Lambda\neq 0$, whereas at $\Lambda=0$, the violation is due to single-particle oscillations.
The violation of the LGI for $\Lambda\neq 0$ implies that macrorealism, which consists of MRPS and NIM, is excluded even for such a large $N$, and thus contradicts what is expected by the GP description; the LGI is never broken for the GP description~\cite{GP}.
In Figs.~\ref{fig:LG_Lambda_N100} and \ref{fig:LG_Lambda_N500}, the LGI is also violated at smaller values of $N$ for $|\Lambda|$ below the critical values near 2. These critical values coincide with the thresholds for the onset of self-trapping for respective $N$.

The self-trapping prevents the system from coherent oscillation and yields $C_{12}=C_{13}=1$ for any interval $\tau$ (see SM). Then $C_{12}$ and $C_{13}$ cancel out and $K=C_{23}\le 1$. Thus, the LGI is nonviolated when self-trapping occurs for $|\Lambda|\geq\Lambda_{\rm c}(N)$.


%

Figure~\ref{fig:LGmax_U} shows the maximal violations of the LGI for particle numbers corresponding to Figs.~\ref{fig:LG_Lambda_N100}\textendash\ref{fig:LG_Lambda_N2000}.
It indicates that the LGI is no longer violated, regardless of $\tau$, once $\Lambda$ exceeds the critical value $\Lambda_{\rm c}(N)$ for each $N$.
Notably, this critical value approaches, but remains below, the threshold for the onset of self-trapping $\Lambda_{\rm st}^{\rm GP}=2$ as $N$ increases.
Remarkably, the violation of the LGI becomes more significant as $N$ increases for fixed $\Lambda$, as shown in Figs.~\ref{fig:LGmax_U} and~\ref{fig:LGmax_N} (also see SM).
 

Z\'arate {\it et al}. investigated the violation of the LGI for two weakly coupled BECs in a double-well potential and found a maximal violation of the LGI 1.5~\cite{Laura}. This is attributed to their presumption of the two-state approximation that results in two-outcome measurements~\cite{Budroni_2013} (also see SM). In contrast, Fig.~\ref{fig:LGmax_U} shows an LG violation exceeding 2, consistent with the result in Ref.~\cite{Budroni}, due to the use of multiple projectors $\ket{n_{\rm L}}\ket{n_{\rm R}}$ and coarse graining of the results by the sign ($\hat S_z$).

Figures~\ref{fig:LG_Lambda_N500} and \ref{fig:LG_Lambda_N2000} show that the period of oscillation of $K$ increases and its amplitude decreases as $\Lambda$ approaches $\Lambda_{\rm c}(N)$ from below. It is due to the softening of the Josephson oscillation associated with the transition from the Josephson regime to the self-trapping regime \cite{Pezze,Farolfi}.
\par
The LGI becomes nonviolated for $\tau$ greater than a characteristic value when $0<\Lambda<\Lambda_{\rm c}(N)$. This characteristic value of $\tau$ becomes smaller as $\Lambda$ increases for fixed $N$ and becomes larger as $N$ increases for fixed $\Lambda$, as shown in Figs.~\ref{fig:LG_Lambda_N100}\textendash\ref{fig:LG_Lambda_N2000}.
This behavior is due to the damping of Josephson oscillation, which destroys the correlation in time.
The Josephson oscillation indeed decays after a characteristic time, which becomes smaller as the interaction strength increases, as shown in Figs.~\ref{fig:te_Lambda1}\textendash\ref{fig:te_Lambda1.9}, but it becomes longer as $N$ increases due to suppression of the fluctuation of the phase difference of two BECs.
\par
The LGI is found to be violated symmetrically between repulsive ($\Lambda>0$) and attractive ($\Lambda<0$) on-site interaction in Figs.~\ref{fig:LG_Lambda_N100}\textendash\ref{fig:LG_Lambda_N2000} since the Hamiltonian with a repulsive interaction can be mapped to the one with an attractive interaction as $\hat R_z(\pi)\hat H_{\rm BJJ}(\Lambda)\hat R_z^\dagger(\pi)=-\hat H_{\rm BJJ}(-\Lambda)$, where $\hat R_z(\theta)=\exp\qty(-i\theta\hat S_z)$. From this relationship, we obtain
\begin{align}
\label{prob_sym}
P(n_{\rm L},\Lambda,t)=P(n_{\rm L},-\Lambda,t),
\end{align}
where $P(n_{\rm L},\Lambda,t)$ denotes the probability of $n_{\rm L}$ bosons on the left-well at time $t$ for $\Lambda$ (see SM). This symmetric relationship between repulsive and attractive on-site interaction results in the symmetric evolution of $K$ between them in Figs.~\ref{fig:LG_Lambda_N100}\textendash\ref{fig:LG_Lambda_N2000}.

\textit{Effect of the projective measurements on the LGI}.
The violations of the LGI in Figs.~\ref{fig:LG_Lambda_N100}\textendash\ref{fig:LG_Lambda_N2000} arise from the absence of either MRPS, NIM, or both. Although it is not possible to conclusively determine whether the violation is specifically caused by one over the other, the violation of the LGI necessarily implies that at least one of these assumptions does not hold. Here, we examine the impact of projective measurements to investigate whether the LGI violation indicates a breakdown of NIM.

\begin{figure}[t!]
\centering
\begin{subfigure}[t]{0.235\textwidth}
\includegraphics[width=\textwidth]{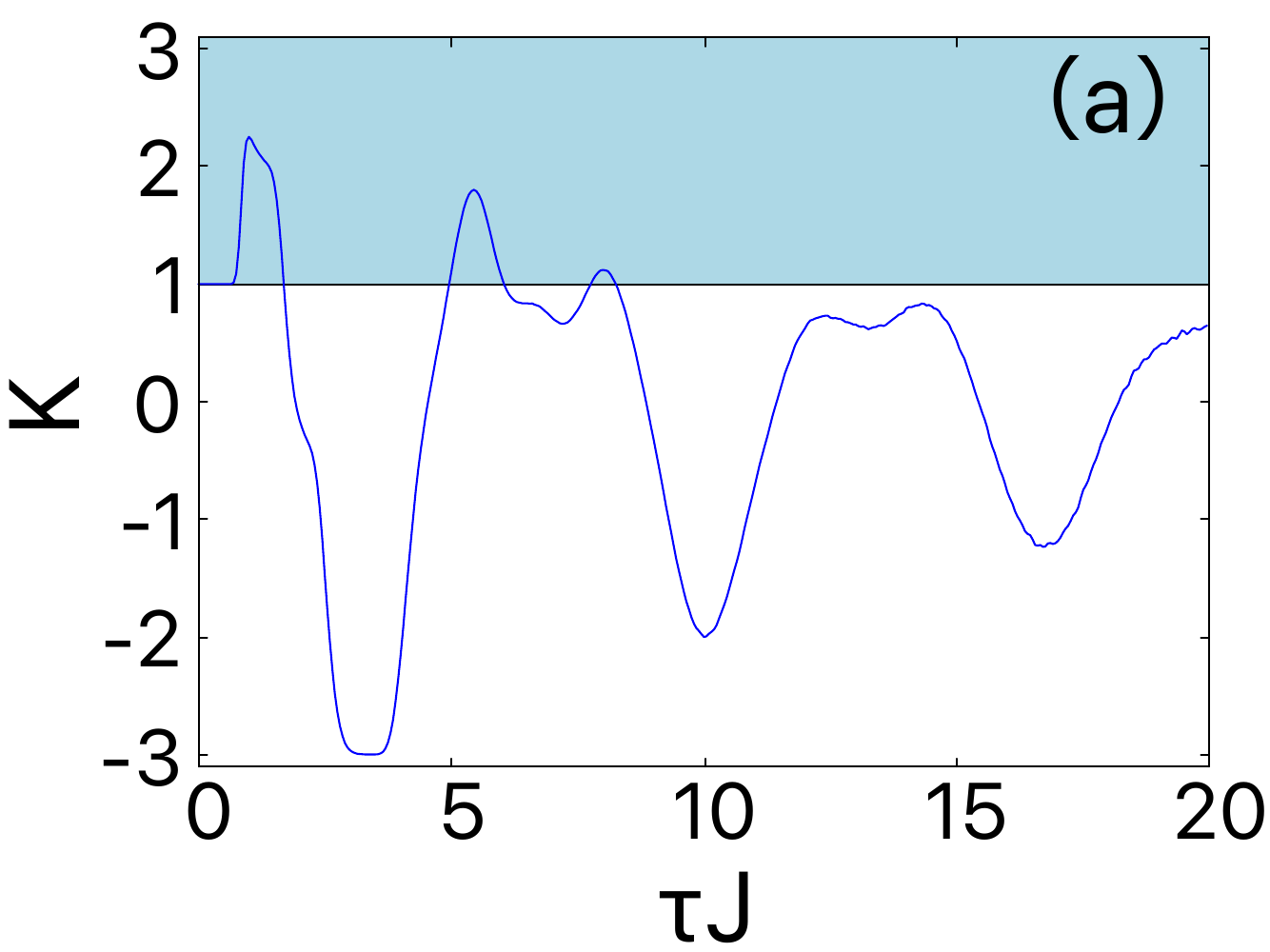}
\phantomsubcaption
\label{fig:LG_col} 
\end{subfigure}
\begin{subfigure}[t]{0.235\textwidth}
\includegraphics[width=\textwidth]{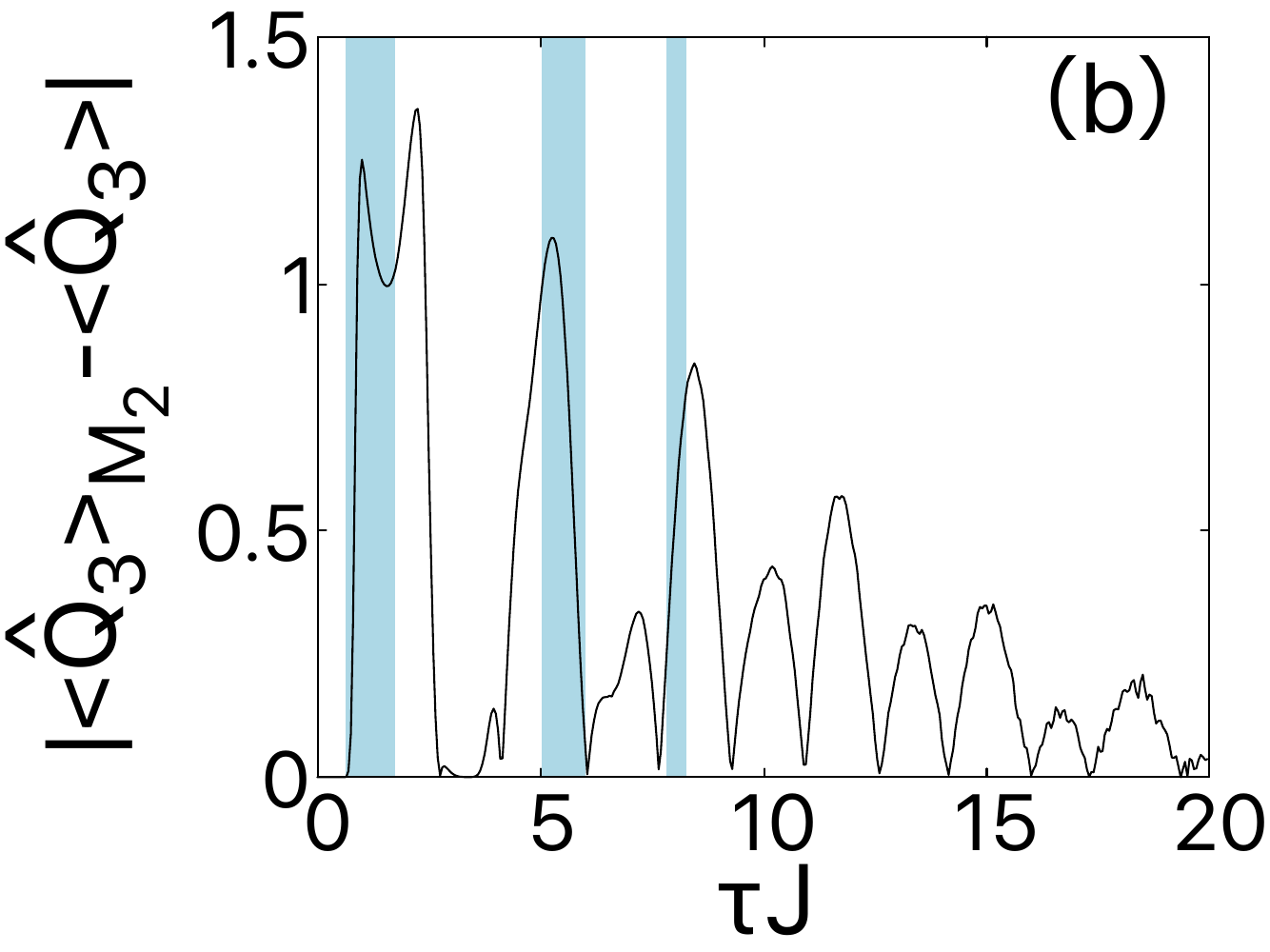}
\phantomsubcaption
\label{fig:collapse} 
\end{subfigure}
\caption{
Effect of the projective measurements at time $t_2$.
(a) Plot of $K$ as a function of $\tau J$. The upper blue region represents a violation of the LGI in Eq.~(\ref{LG}). 
(b) $|\langle Q_3\rangle_{{\rm M}_2}-\langle Q_3\rangle|$ as a function of $\tau J$. 
The blue region indicates $\tau J$ at which the LGI is violated in (a).
The parameters used in (a) and (b) are $N=100$ and $\Lambda=1$.
}
\label{fig:LG_inv} 
\end{figure}

The effect of projective measurements appears explicitly in Fig.~\ref{fig:collapse} in the comparison between $\langle Q_3\rangle_{{\rm M}_2}$, where the state is assumed to be collapsed due to measurements at $t_2$, and $\langle Q_3\rangle$, where measurements are not made at $t_2$. The condition $|\langle Q_3\rangle_{{\rm M}_2}-\langle Q_3\rangle|\neq0$ implies a breakdown of NIM, indicating that the dynamics after time $t_2$ evolves differently depending on whether a measurement is made at $t_2$ or not. A comparison between Figs.~\ref{fig:LG_col} and \ref{fig:collapse} shows that whenever the LGI is violated, $|\langle Q_3\rangle_{{\rm M}_2}-\langle Q_3\rangle|$ is nonzero. Moreover, the greater the violation of the LGI, the larger is the value of $|\langle Q_3\rangle_{{\rm M}_2}-\langle Q_3\rangle|$ (also see SM). 
Note that 
$|\langle Q_3\rangle_{{\rm M}_2}-\langle Q_3\rangle|\neq0$ does not necessarily lead to violation of the LGI.
In Fig.~\ref{fig:LG_col}, the oscillation of $K$ decays, and the LGI becomes nonviolated. Accordingly, the effect of projective measurements disappears in Fig.~\ref{fig:collapse}.

\textit{Observability of violation of the LGI}.
\begin{figure}[t!]
\centering
\begin{subfigure}[t]{0.23\textwidth}
\includegraphics[width=\textwidth]{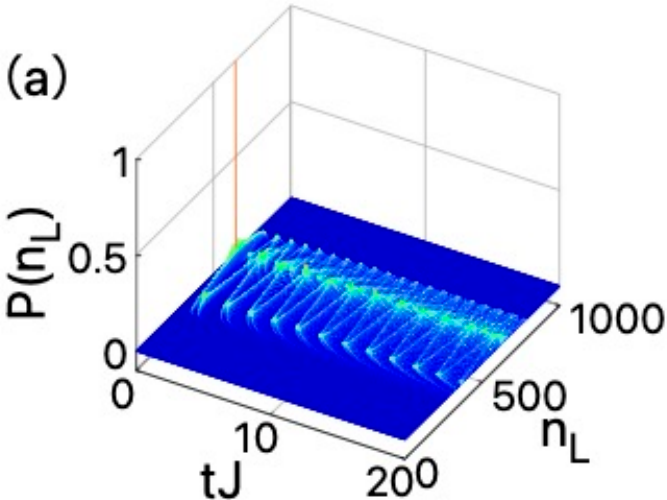}
\phantomsubcaption
\label{fig:Albiez_jo} 
\end{subfigure}
\begin{subfigure}[t]{0.23\textwidth}
\includegraphics[width=\textwidth]{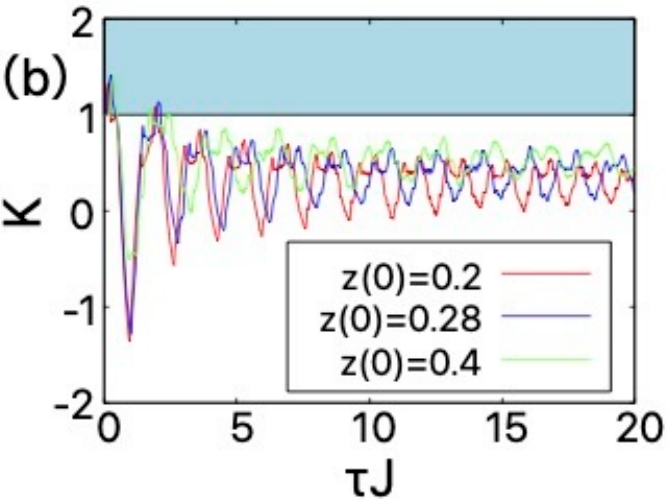}
\phantomsubcaption
\label{fig:LG_N1000_U15_z} 
\end{subfigure}
\caption{(a) Time evolutions of $P(n_{\rm L})$ for $z(0)=0.28$. 
(b) Violations of the LGI for $z(0)=0.2$ (the red line), 0.28 (the blue line), and 0.4 (the green line). 
These parameters are well below the critical value of macroscopic quantum self-trapping. 
In (a) and (b), the parameters are $N=1000$ and $\Lambda=15$.}
\end{figure}
We numerically test the LGI for nearly the same parameters for which Josephson oscillations were observed in Ref.~\cite{Albiez}. Here, we set $\Lambda=15$ and $N=1000$.
The LGI is shown to be violated for realistic parameters in Fig.~\ref{fig:LG_N1000_U15_z}, in which the evolution of $K$  is plotted as a function of $\tau$ for several initial population imbalances $z(0)=(n_{\rm L}(0)-n_{\rm R}(0))/N$.
$z(0)$ needs to be smaller than the critical value for self-trapping $z_{\rm st}(0)\simeq 0.5$~\cite{Raghavan}.
$z(0)=0.28$ is expected to be more desirable for the demonstration of violation of the LGI.
Figure~\ref{fig:Albiez_jo} shows the time evolution of $P(n_{\rm L})$ for $z(0)=0.28$. It exhibits the tunneling of atoms between the two wells, which indeed leads to the violation of the LGI.

\textit{Single-particle regime}.
Next, we study violation of the LGI in the absence of an on-site interaction. 
In this case the Hamiltonian is simply given by $\hat H_{\rm BJJ}=-J\hat S_x$.
For this model, the time-dependent wave function can be obtained analytically.
The wave function at time $t$ evolved from the initial state $|N-l\rangle_{\rm L}|l\rangle_{\rm R}$ is given by (see SM)
\begin{align}\label{wave_initial}
\ket{\Psi(t)}=&\frac{1}{\sqrt{(N-l)!}}\frac{1}{\sqrt{l!}}\left(\hat a^\dagger_{\rm L}\cos \frac{J}{2}t-i\hat a^\dagger_{\rm R}\sin \frac{J}{2}t\right)^{N-l}\notag\\
&\times \left(\hat a^\dagger_{\rm R}\cos \frac{J}{2}t-i\hat a^\dagger_{\rm L}\sin \frac{J}{2}t\right)^{l}|\mathrm{vac}\rangle.
\end{align}

\begin{figure}[t!]
\centering
\begin{subfigure}[t]{0.23\textwidth}
\includegraphics[width=\textwidth]{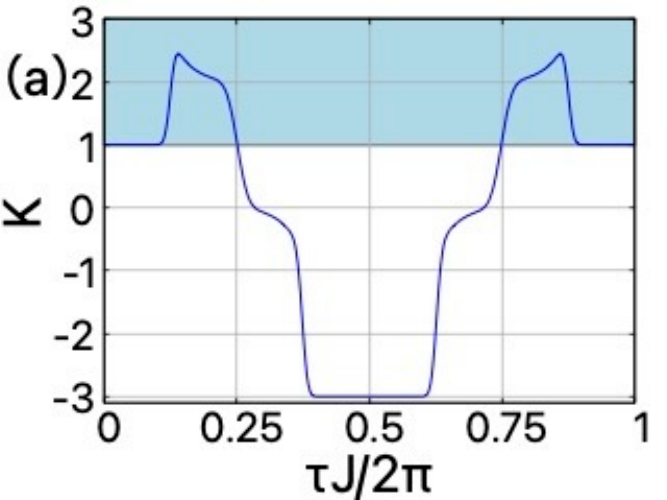}
\phantomsubcaption
\label{fig:LG_N100_U0}
\end{subfigure}
\begin{subfigure}[t]{0.23\textwidth}
\includegraphics[width=\textwidth]{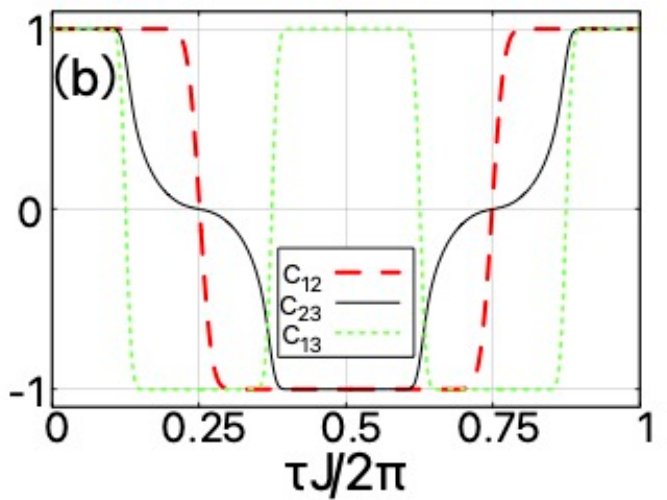}
\phantomsubcaption
\label{fig:TTC_N100_U0}
\end{subfigure}
\begin{subfigure}[t]{0.235\textwidth}
\includegraphics[width=\textwidth]{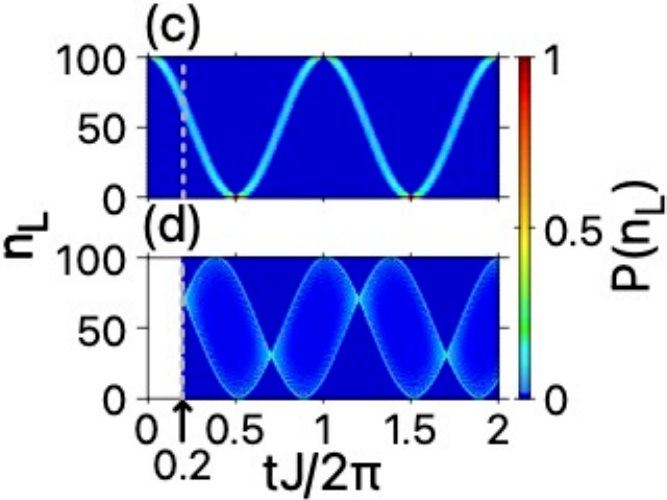}
\phantomsubcaption
\label{fig:coherent_te}
\phantomsubcaption
\label{fig:te_t0.2_M}
\end{subfigure}
\begin{subfigure}[t]{0.235\textwidth}
\includegraphics[width=\textwidth]{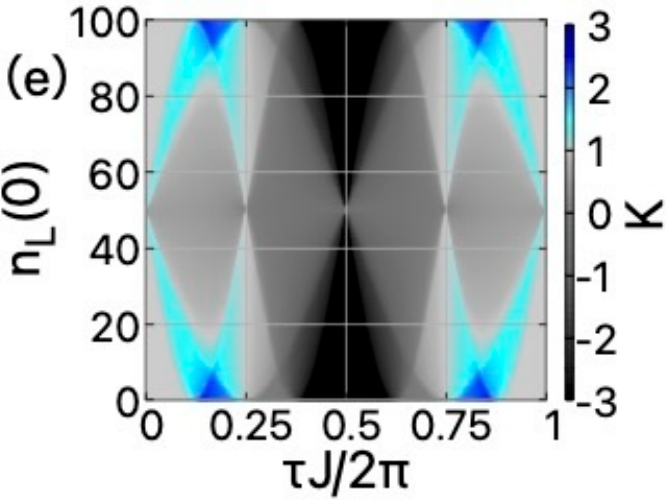}
\phantomsubcaption
\label{fig:LG_initial}
\end{subfigure}
\caption{(a) Violation of the LGI. (b) Three correlation functions $C_{12}$ (the red dashed line), $C_{23}$ (the black solid line), and $C_{13}$ (the green dotted line) as functions of $\tau$. 
(c)/(d) Time evolution of $P(n_{\rm L})$ in the absence/presence of the measurement at $tJ/2\pi=0.2$ (represented by the gray dotted line). $n_{\rm L}=70$ at $tJ/2\pi=0.2$ in (d). (e) Color map of $K$ for all different initial occupation numbers in the left well $n_{\rm L}(0)$. 
The violation of the LGI is symmetric under the reflection with respect to the line $n_{\rm L}(0)=50$.
The blue color in (e) shows the part where the LGI is violated. In (a)\textendash(e), the parameters are $N=100$ and $\Lambda=0$.}
\label{fig:LG_U0} 
\end{figure} 

The first (second) line of Eq.~(\ref{wave_initial}) represents the coherent oscillation of the bosons initially located in the left (right) well. These bosons flow in opposite directions. The former represents bosons initially tunneling from the left well to the right well, as shown in Fig.~\ref{fig:coherent_te}, while the latter represents those initially tunneling from the right well to the left well.
As shown in Fig.~\ref{fig:coherent_te}, the peak of 
the occupation probability coherently oscillates between $n_{\rm L}=0$ and $n_{\rm L}=N$ without damping.
Due to this coherent oscillation the LGI is violated, as shown in Fig.~\ref{fig:LG_N100_U0}. 

The violation of the LGI in Fig.~\ref{fig:LG_N100_U0} closely resembles the one presented in Fig. 2(a) in Ref.~\cite{Lambert}. This is due to the special coarse graining of the results as the sign ($S_z$). Our work represents a concrete realization of the proposal of Lambert {\it et al}.
for the coarse-grained implementation of LGI violations.

Focusing on the first violation in Fig.~\ref{fig:LG_N100_U0}, this behavior can be explained as follows: 
The LGI holds until $C_{13}$ starts to decrease, since $C_{12}=C_{13}=C_{23}=1$. 
The LGI begins to be first violated when $C_{13}$ starts decreasing at $\tau J/2\pi\simeq 0.125$. Since $C_{13}$ abruptly becomes $-1$ while $C_{12}=1$ and $C_{23}$ is still positive, the LGI becomes violated quite abruptly. The value of the peak of $K$ becomes larger with the increase of $N$, due to a more abrupt decrease of $C_{13}$ (see SM). 

The persistence of this violation is attributed to the bosons tunneling backwards in Fig.~\ref{fig:te_t0.2_M} induced by the measurement, which corresponds to the second line of Eq.~(\ref{wave_initial}). In the case of $C_{23}$ the bosons tunneling backwards impede the decrease of $C_{23}$, while such an effect is absent in the case of $C_{13}$, where no measurement is performed at $t_2$ [see Fig.~\ref{fig:coherent_te}].
The LGI again begins to hold approximately when $C_{12}$ becomes zero at $\tau J/2\pi\simeq 0.25$ [see Fig.~\ref{fig:TTC_N100_U0}].

We also test the LGI for all different initial occupation numbers of the left well $n_{\rm L}(0)$ with $N=100$. In Fig. \ref{fig:LG_initial}, it can be seen that the peak value of $K$ increases with a larger initial population imbalance (see SM).

\textit{Experimental realization}.
The experimental setup in Ref.~\cite{Albiez} provides a concrete realization of the theoretical predictions in our paper. Phase contrast imaging (PCI) is a promising candidate for implementing our measurement scheme as a nondestructive measurement of a BEC in a double-well potential~\cite{Andrews, Bradley, Higbie, Ilo-Okeke}. PCI can perform measurements in the number basis $\ket{n_{\rm L}}\ket{n_{\rm R}}$, enabling the measurement of the population in each well nondestructively without atom loss. This approach allows for the acquisition of each correlation function.

\textit{Conclusion and discussion}.
In this Letter, we study violations of the LGI for bosons in a double-well potential. 
We find that the LGI is violated due to Josephson oscillations in the weak-coupling regime, while it remains unviolated in the strong-coupling regime for $\Lambda$ beyond the critical $\Lambda_{\rm c}$, which depends on the particle number $N$. $\Lambda_{\rm c}$ approaches, but remains below, the critical threshold for the onset of self-trapping, $\Lambda_{\rm st}^{\rm GP}=2$, as derived from semiclassical analysis.
It indicates that the nonviolation of the LGI is attributed to the self-trapping phenomena. 
Remarkably, we observe that the violation of the LGI becomes increasingly significant as the particle number increases.
We also find that, in this system, an on-site interaction induces a dephasing effect that destroys temporal correlations, resulting in the LGI no longer being violated afterwards.
Furthermore, we have revealed that the LGI is violated for nearly the same parameters for which the Josephson oscillations are observed in Ref.~\cite{Albiez}, thus indicating the presence of macroscopic quantum coherence in the semiclassical regime.



\textit{Acknowledgments}.
We thank anonymous reviewers for insightful comments. We thank G. Kimura, M. Kunimi, and T. Nikuni for fruitful discussions. S.T. thanks F. Dalfovo, G. Ferrari, and A. Recati for useful discussions. R.Y. is sup- ported by the Japanese Society for the Promotion of Science Grant-in-Aid for Scientific Research (KAKENHI, Grants No. 19K14616 and No. 20H01838) and of the WPI program “Sustainability with Knotted Chiral Meta Matter (SKCM2)” at Hiroshima University. S.T. is supported by the Japan Society for the Promotion of Science Grant-in-Aid for Scientific Research (KAKENHI Grant No. 19K03691).

\end{document}


\title{Supplemental Materials for Violation of the Leggett-Garg inequality for dynamics of a Bose-Einstein condensate in a double-well potential}

\author{Tsubasa Sakamoto$^1$, Ryosuke Yoshii$^{2,3}$, Shunji Tsuchiya$^1$}
\affiliation{$^1$Department of Physics, Chuo University, 1-13-27 Kasuga, Bunkyo-ku, Tokyo 112-8551, Japan\\
$^2$Center for Liberal Arts and Sciences, Sanyo-Onoda City University, 1-1-1 Daigaku-Dori, Sanyo-Onoda, Yamaguchi 756-0884, Japan\\$^3$International Institute for Sustainability with Knotted Chiral Meta Matter (WPI-SKCM2), Hiroshima University, Higashi-Hiroshima, Hiroshima 739-8526, Japan}

\date{\today}

\maketitle

\section{Time evolution of the correlation functions}\label{appendix:A}

\subsection{Numerical method}
In our tests, we define the dichotomic observable as $\hat Q={\rm sgn}(\hat S_z)$, where ${\rm sgn}(x)=1$ if $x\geq0$, otherwise ${\rm sgn}(x)=-1$, and focus on the identical time interval: $\tau=t_2-t_1=t_3-t_2$. Also note that in our tests we set an initial state at time $t_1$. Thus, $Q_1$ is fixed by the preparation of the initial state rather than by performing measurements on the system. $\ev{Q_1Q_2}$ and $\ev{Q_1Q_3}$ are then given by
\begin{gather}\label{Q1Q2}
    \ev{Q_1Q_2}={\rm sgn}(n_{\rm L,1}-n_{\rm R,1})\sum_{n_{\rm L,2}}{\rm sgn}(n_{\rm L,2}-n_{\rm R,2})P(n_{\rm L,2},\tau),\\
    \label{Q1Q3}\ev{Q_1Q_3}={\rm sgn}(n_{\rm L,1}-n_{\rm R,1})\sum_{n_{\rm L,3}}{\rm sgn}(n_{\rm L,3}-n_{\rm R,3})P(n_{\rm L,3}, 2\tau),
\end{gather}
where $n_{{\rm L/R},i}$ denotes the value of $n_{\rm L/R}$ associated with $Q_i$. Thus $\ev{Q_1Q_2}$ and $\ev{Q_1Q_3}$ only require measurements at $t_2$ and $t_3$, respectively. Note that when we calculate $\ev{Q_1Q_3}$ we don't perform measurement at $t_2$.  Also note that for $\ev{Q_1Q_2}$ and $\ev{Q_1Q_3}$ there is no concern about the disturbance of measurements because we are no longer interested in the subsequent dynamics.

$\ev{Q_2Q_3}$ is more complicated than the two other correlation functions and given by  
\begin{gather}\label{Q2Q3}
    \ev{Q_2Q_3}=\sum_{n_{\rm L,2}}{\rm sgn}(n_{\rm L,2}-n_{\rm R,2})P(n_{\rm L,2},\tau)\sum_{n_{\rm L,3}}{\rm sgn}(n_{\rm L,3}-n_{\rm R,3})P(n_{\rm L,3},2\tau|n_{\rm L,2},\tau),
\end{gather}
where $P(n_{\rm L,3},2\tau|n_{\rm L,2},\tau)$ denotes the probability of obtaining $\ket{n_{\rm L,3}}_{\rm L}\ket{N-n_{\rm L,3}}_{\rm R}$ at $t_3$ after obtaining $\ket{n_{\rm L,2}}_{\rm L}\ket{N-n_{\rm L,2}}_{\rm R}$ at $t_2$.
Unlike $\ev{Q_1Q_2}$ and $\ev{Q_1Q_3}$, $\ev{Q_2Q_3}$ requires measurements at two times. Crucially, the first measurements at $t_2$ concern about the disturbance on the subsequent dynamics, while the last measurements at $t_3$ don't. One wishes to observe $n_{\rm L,2}$ without disturbing the subsequent dynamics of the system. $n_{\rm L,2}$, however, is a fluctuating quantity through the time-evolution and thus we have no choice but to collapse the wave function at $t_2$ to obtain the definite value of $n_{\rm L,2}$. In our numerical method, to calculate $P(n_{\rm L,3},2\tau|n_{\rm L,2},\tau)$, we set $\ket{n_{\rm L,2}}_{\rm L}\ket{N-n_{\rm L,2}}_{\rm R}$ at $t_2$ and evolve the wave function from this state and later perform measurements at $t_3$.

\subsection{Interacting bosons}
\begin{figure}[t!]
\includegraphics[width=0.5\textwidth]{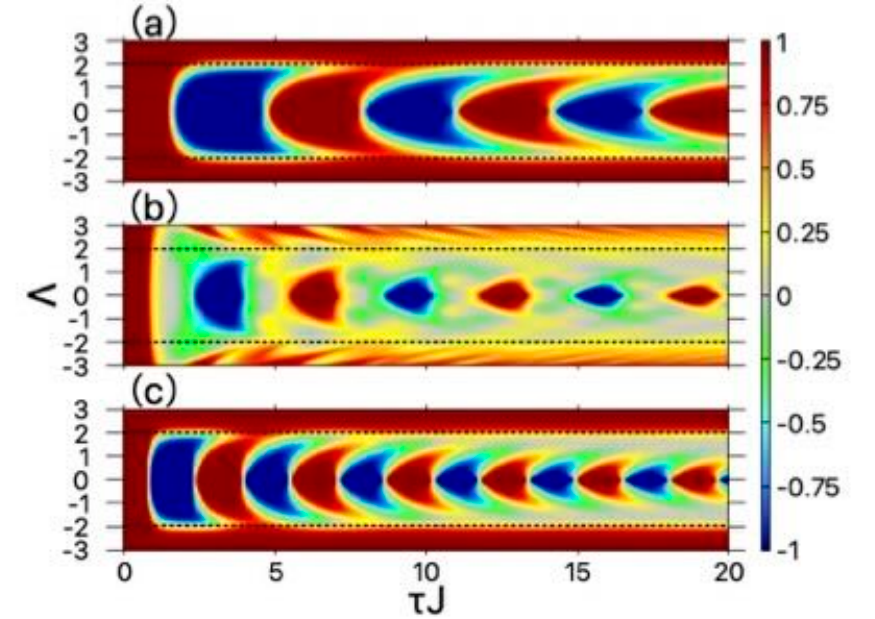}
\caption{
Colormaps of the three correlation functions $C_{12}$ (a), $C_{23}$ (b), and $C_{13}$ (c) for $N=100$. 
The two black dashed lines in respective panels denote the critical value of macroscopic quantum self-trapping $|\Lambda_{\rm c}|=2$.}
\label{fig:selftrapping} 
\end{figure}

We here present the results of the three correlation functions for interacting bosons. 
In Fig.~\ref{fig:selftrapping}, the time evolutions of the correlation functions are plotted as a function of $\tau$ with $n_{\rm L}(0)=N$. 
The figure shows that the correlation functions hardly change outside of the critical lines $\Lambda_{\rm c}=\pm2$. 
These critical lines coincide with the self-trapping threshold evaluated from the semiclassical treatment~\cite{Raghavan}. 

Even inside the critical lines, the amplitude of the correlation functions dumps in time. 
This behavior can be interpreted as the dephasing of the oscillations of the occupation probabilities due to the on-site interaction.

\subsection{Non-interacting bosons}

\begin{figure}[t!]
\centering
\begin{subfigure}[t]{0.3\textwidth}
\includegraphics[width=\textwidth]{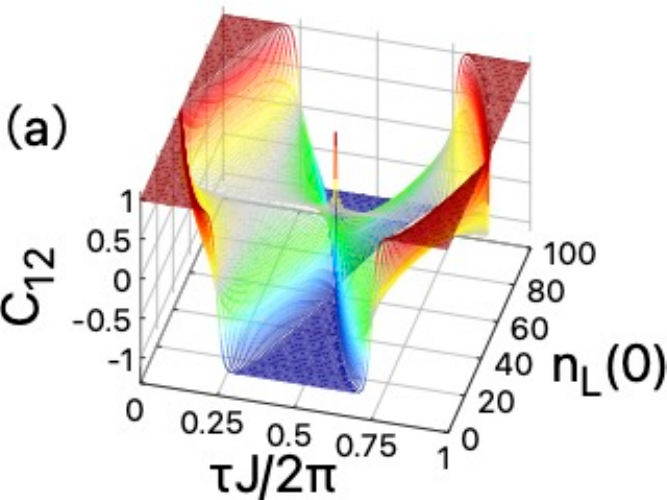}
\phantomsubcaption
\label{fig:LG_col} 
\end{subfigure}
\begin{subfigure}[t]{0.3\textwidth}
\includegraphics[width=\textwidth]{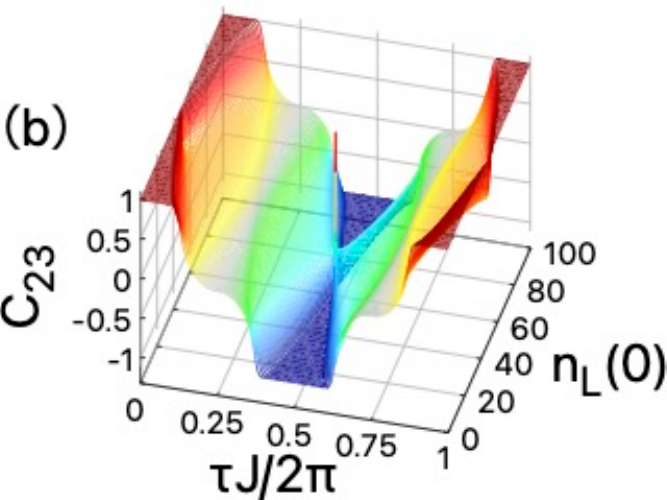}
\phantomsubcaption
\label{fig:collapse} 
\end{subfigure}
\begin{subfigure}[t]{0.3\textwidth}
\includegraphics[width=\textwidth]{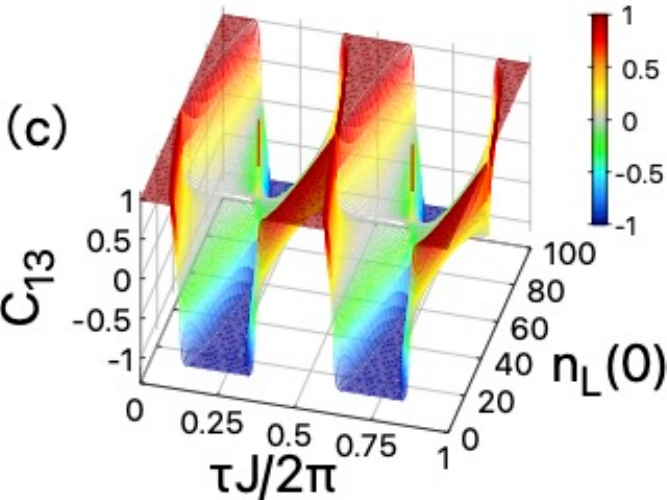}
\phantomsubcaption
\label{fig:collapse} 
\end{subfigure}
\caption{Three correlation functions $C_{12}$ (a), $C_{23}$ (b), and $C_{13}$ (c) as functions of a given interval variable $\tau$ for different $n_{\rm L}(0)$. Here we set $N=100$.}
\label{fig:cols} 
\end{figure}

We now turn off on-site interaction and we consider the dependencies of the three correlation functions on the initial population imbalance.

The Leggett-Garg inequality (LGI) $C_{12}+C_{23}-C_{13}\le 1$ is not violated at the beginning, since all quantities $C_{12},\ C_{23},\ C_{13}$ are unity, as can be seen in Fig.~\ref{fig:cols} and thus $C_{12}+C_{23}-C_{13}=1$. 
In the present case, the characteristic time for the dumping is shortest for $C_{13}$ and thus the LGI is broken once $C_{13}$ starts to decrease.

\subsection{Analytical calculations of the correlation functions in non-interacting case}\label{appendix:D}
Here, we will show the analytical calculations of $C_{12}$, $C_{23}$ and $C_{13}$. First, we will derive the wave function for an arbitral time in the non-interacting case with the double well potential.

The Hamiltonian of the system is given by
\begin{equation}
\hat H=-\frac{J}{2}(\hat a_{\rm L}^\dagger \hat a_{\rm R}+\hat a_{\rm R}^\dagger \hat a_{\rm L}),  
\end{equation}
where $\hat a^\dagger_{\rm L} (\hat a_{\rm L}), \hat a^\dagger_{\rm R}\ (\hat a_{\rm R})$ are the creation (annihilation) operators of the boson in the left and the right well, respectively. 
We consider the initial state $\ket{N-l}_{\rm L}\ket{l}_{\rm R}$. 
By rewriting the initial state in the following form, one can easily find the state at arbitral time:
\begin{equation}
\ket{N-l}_{\rm L}\ket{l}_{\rm R}=\frac{1}{\sqrt{(N-l)!}}\frac{1}{\sqrt{l!}}\left(\hat a^\dagger_{\rm L}\right)^{N-l}\left(\hat a^\dagger_{\rm R}\right)^{l}|\mathrm{vac}\rangle, 
\end{equation}
where $|\rm{vac}\rangle$ is the empty state,
In the Heisenberg picture, the time evolution of the operators can be calculated as follows:
\begin{align}
&\frac{d}{dt}\hat a^{\dagger}_{\rm L}(t)=i[\hat H,\hat a^\dagger_{\rm L}(t)]=ie^{i\hat Ht}[\hat H,\hat a^\dagger_{\rm L}]e^{-i\hat Ht}=-\frac{J}{2}i\hat a^\dagger_{\rm R}(t),\\ 
&\frac{d}{dt}\hat a^{\dagger}_{\rm R}(t)=i[\hat H,\hat a^\dagger_{\rm R}(t)]=ie^{i\hat Ht}[\hat H,\hat a^\dagger_{\rm R}]e^{-i\hat Ht}=-\frac{J}{2}i\hat a^\dagger_{\rm L}(t). 
\end{align}
These equations can be recast into the matrix form as 
\begin{align}
\frac{d}{dt}\left(\begin{array}{c}
\hat a^{\dagger}_{\rm L}(t)\\
\hat a^{\dagger}_{\rm R}(t)
\end{array}\right)
=-i\frac{J}{2}\left(\begin{array}{cc}
0&1\\
1&0
\end{array}\right)
\left(\begin{array}{c}
\hat a^{\dagger}_{\rm L}(t)\\
\hat a^{\dagger}_{\rm R}(t)
\end{array}\right).
\end{align}
This can be easily solved as 
\begin{align}
\left(\begin{array}{c}
\hat a^{\dagger}_{\rm L}(t)\\
\hat a^{\dagger}_{\rm R}(t)
\end{array}\right)
=\left(\begin{array}{cc}
\cos \frac{J}{2}t&-i\sin \frac{J}{2}t\\
-i\sin \frac{J}{2}t&\cos \frac{J}{2}t
\end{array}\right)
\left(\begin{array}{c}
\hat a^{\dagger}_{\rm L}(0)\\
\hat a^{\dagger}_{\rm R}(0)
\end{array}\right).
\end{align}
Thus the state prepared as $\ket{N-l}_{\rm L}\ket{l}_{\rm R}$ at $t=0$ evolves in time as 
\begin{align}\label{wave_arb}
|\Psi(l,t)\rangle=&\frac{1}{\sqrt{(N-l)!}}\frac{1}{\sqrt{l!}}\left(\hat a^\dagger_{\rm L}(t)\right)^{N-l}\left(\hat a^\dagger_{\rm R}(t)\right)^{l}|\mathrm{vac}\rangle\notag\\
=&\frac{1}{\sqrt{(N-l)!}}\frac{1}{\sqrt{l!}}\left(\hat a^\dagger_{\rm L}\cos \frac{J}{2}t-i\hat a^\dagger_{\rm R}\sin \frac{J}{2}t\right)^{N-l}\notag\\
&\times \left(\hat a^\dagger_{\rm R}\cos \frac{J}{2}t-i\hat a^\dagger_{\rm L}\sin \frac{J}{2}t\right)^{l}|\mathrm{vac}\rangle. 
\end{align}


Eq.~(\ref{wave_arb}) can be rewritten as follows;
\begin{align}
\ket{\Psi(l,t)}=&\frac1{\sqrt{(N-l)!}\sqrt{l!}}\sum_{\mu=0}^{N-l}\comb{N-l}{\mu}\qty(\aL\cos\frac J2t)^{N-l-\mu}\qty(-i\aR\sin\frac J2t)^{\mu}\\
&\times\sum_{\nu=0}^l\comb{l}{\nu}\qty(\aR\cos\frac J2t)^{\nu}\qty(-i\aL\sin\frac J2t)^{l-\nu}\ket{{\rm vac}}\\
=&\frac1{\sqrt{(N-l)!}\sqrt{l!}}\sum_{\mu=0}^{N-l}\sum_{\nu=0}^l\comb{N-l}{\mu}\comb{l}{\nu}\qty(\cos\frac J2t)^{N-l-\mu+\nu}\qty(-i\sin\frac J2t)^{l+\mu-\nu}\\
&\times\sqrt{(N-k)!}\sqrt{k!}\ket{N-k}_{\rm L}\ket{k}_{\rm R},
\end{align}
where $k=\mu+\nu$. Thus, the occupation probabilty of the right well in the case of $\ket{\Psi(0)}=\ket{N-l}_{\rm L}\ket{l}_{\rm R}$ is obtained as follows,
\begin{align}\label{P_arb}
P(l,n_{\rm R},t)=&\frac{(N-n_{\rm R})!n_{\rm R}!}{(N-l)!l!}\qty|\sum_{\{\mu,\nu|\mu+\nu=n_{\rm R}\}}\comb{N-l}{\mu}\comb{l}{\nu}\qty(\cos\frac J2t)^{N-l-\mu+\nu}\qty(-i\sin\frac J2t)^{l+\mu-\nu}|^2.
\end{align}
From Eq.~(\ref{P_arb}), we obtain the three correlation functions as follows:
\begin{align}
C_{12}=&\sum_{n_{{\rm R},2}=0}^{\floor{N/2}}P(0,n_{{\rm R},2},\tau)-\sum_{n_{{\rm R},2}=\floor{N/2}+1}^{N}P(0,n_{{\rm R},2},\tau),\\
C_{23}=&\sum_{n_{{\rm R},2}=0}^{\floor{N/2}}P(0,n_{{\rm R},2},\tau)\qty(\sum_{n_{{\rm R},3}=0}^{\floor{N/2}}P(n_{{\rm R},2},n_{{\rm R},3},\tau)-\sum_{n_{{\rm R},3}=\floor{N/2}+1}^{N}P(n_{{\rm R},2},n_{{\rm R},3},\tau))\nonumber\\
&-\sum_{n_{{\rm R},2}=\floor{N/2}+1}^{N}P(0,n_{{\rm R},2},\tau)\qty(\sum_{n_{{\rm R},3}=0}^{\floor{N/2}}P(n_{{\rm R},2},n_{{\rm R},3},\tau)-\sum_{n_{{\rm R},3}=\floor{N/2}+1}^{N}P(n_{{\rm R},2},n_{{\rm R},3},\tau)),\\
C_{13}=&\sum_{n_{{\rm R},3}=0}^{\floor{N/2}}P(0,n_{{\rm R},3},2\tau)-\sum_{n_{{\rm R},3}=\floor{N/2}+1}^{N}P(0,n_{{\rm R},3},2\tau).
\end{align}

In the present setup, $C_{12}$ and $C_{13}$ reduces to the simple form since  
\begin{align}
P(0,n_{\rm R},t)&=\frac{(N-n_{\rm R})!n_{\rm R}!}{N!}\qty|\comb{N}{n_{\rm R}}\qty(\cos\frac J2t)^{N-n_{\rm R}}\qty(-i\sin\frac J2t)^{n_{\rm R}}|^2\notag\\
&=\comb{N}{n_{\rm R}}\qty(\cos^2\frac J2t)^{N-n_{\rm R}}\qty(\sin^2\frac J2t)^{n_{\rm R}}.
\end{align}
The summation appearing in $C_{12}$ or $C_{13}$ becomes 
\begin{align}
S_1=\sum_{n_{{\rm R},2}=0}^{\floor{N/2}}P(0,n_{{\rm R},2},\tau)
=c(\tau)^N+N c(\tau)^{N-1}s(\tau)+\cdots \frac{N!}{(N-\floor{N/2})!\floor{N/2}!}c(\tau)^{N-\floor{N/2}}s(\tau)^{\floor{N/2}},
\end{align}
where we define $c(\tau)=\cos^2\frac J2\tau$ and $s(\tau)=\sin^2\frac J2\tau$. 
The differentiation of $S_1$ with respect to $\tau$ becomes 
\begin{align}
\frac{d}{d\tau}S_1=\frac{dc(\tau)}{d\tau}\frac{N!}{(N-\floor{N/2})!\floor{N/2}!}(N-\floor{N/2})c(\tau)^{N-\floor{N/2}-1}s(\tau)^{\floor{N/2}},
\end{align}
where we use the relation $\frac{d}{d\tau}c(\tau)=-\frac{d}{d\tau}s(\tau)=-J\sin\frac J2\tau\cos\frac J2\tau$.
Thus we obtain 
\begin{align}
\frac{d}{d\tau}S_1&=-J\frac{N!}{(N-\floor{N/2}-1)!\floor{N/2}!}\qty(\cos\frac J2\tau)^{2N-2\floor{N/2}-1}\qty(\sin\frac J2\tau)^{2\floor{N/2}+1}\nonumber\\
&=-J\frac{N!}{(N-\floor{N/2}-1)!\floor{N/2}!}\qty(\cos\frac J2\tau)^{2N-4\floor{N/2}-2}\qty(\frac{1}{2}\sin J\tau)^{2\floor{N/2}+1}.
\end{align}
By using the Stirling's formula $n!\simeq n^{n+1/2}/e^{-n+1}$, we can further evaluate as 
\begin{align}
\frac{d}{d\tau}S_1&\simeq -J\frac{N^{N+1/2}}{e^{-N+1}}\frac{e^{-(N-\floor{N/2}-1)+1}}{(N-\floor{N/2}-1)^{(N-\floor{N/2}-1)+1/2}}\frac{e^{-\floor{N/2}+1}}{\floor{N/2}^{\floor{N/2}+1/2}}\qty(\cos\frac J2\tau)^{2N-4\floor{N/2}-2}\qty(\frac{1}{2}\sin J\tau)^{2\floor{N/2}+1}.
\end{align}
For sufficiently large $N\gg 1$, $N-\floor{N/2}-1\simeq \floor{N/2}\simeq N/2$ and thus we obtain
\begin{align}
\frac{d}{d\tau}S_1&\simeq -J\frac{N^{N+1/2}}{(N/2)^N}\qty(\cos\frac J2\tau)^{2N-4\floor{N/2}-2}\qty(\frac{1}{2}\sin J\tau)^{2\floor{N/2}+1}.
\end{align}
From this expression, we find that $\frac{d}{d\tau}S_1\simeq 0$ from the initial time to the characteristic time $\tau_0$, which satisfies the following. 
\begin{align}
N^{{1}/{2({2\floor{N/2}+1})}}\sin J\tau_0 \simeq  1. 
\end{align}
Since $C_{12}$ and $C_{13}$ are $1$ at $\tau=0$, $C_{12}$ and $C_{13}$ deviates from $1$ after $\tau_0$ and $2\tau_0$, respectively.

\begin{figure}[t!]
\begin{subfigure}[t]{0.3\textwidth}
\includegraphics[width=\textwidth]{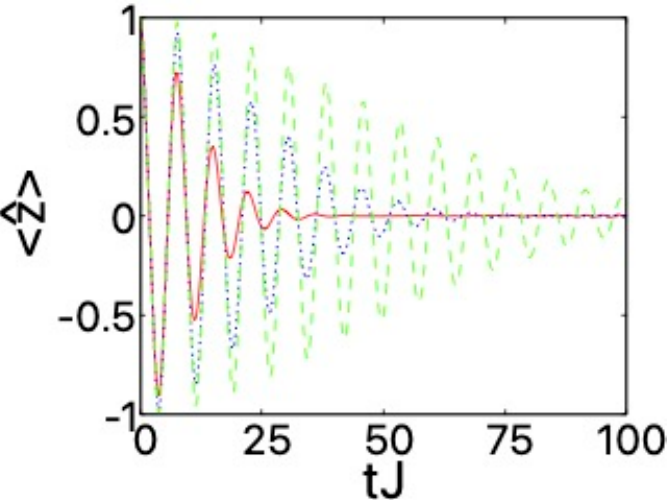}
\end{subfigure}
\caption{Damped oscillations of the population imbalance between the two wells. The red solid line is $N=100$, the blue dotted line $N=500$, and the green dashed line $N=2000$. $\langle \hat z\rangle$ denotes the expectation value of the population imbalance: $\hat z=(\hat n_{\rm L}-\hat n_{\rm R})/N$. Here $\Lambda=1.5$.}
\label{fig:z_U1.5_N} 
\end{figure}

\section{Effect of particle number on dynamics}\label{appendix:B}

In this section, we delve into the effect of particle number on the oscillations of the population imbalance between two wells.

When influenced by on-site interaction, the dephasing phenomenon of the oscillations of population imbalance becomes apparent. As discussed in Ref.~\cite{Milburn}, the fluctuation of the phase difference between the two wells is considered to result in the dephasing effect. Such a fluctuating quantity is assumed to be small enough to be ignored in the semi-classical approach, whilst the effect of the fluctuation on dynamics is taken into account in the Schr\"odinger equation. This is the reason why, unlike what is expected by the GP equation, Josephson oscillations are eventually dumped. As clarified in Ref.~\cite{Gajda}, quantum fluctuation in the left (right) well becomes smaller with the increase of $n_{\rm L}$ ($n_{\rm R}$).  
This is the reason why, as $N$ increases, the coherent Josephson oscillation endures for longer, as shown in Fig.~\ref{fig:z_U1.5_N}.

\section{Symmetry between repulsive and attractive on-site interaction}\label{appendix:C}

Here we prove that for any initial states the evolution of the occupation probability $P(n_{\rm L})$ is symmetric between repulsive $(\Lambda>0)$ and attractive $(\Lambda<0)$ on-site interaction. 

By rotating the Hamiltonian $\hat H_{\rm BJJ}$ about the z axis, we have
\begin{align}
\label{BJJ_rot_z}\hat R_z(\pi)\hat H_{\rm BJJ}(\Lambda)\hat R_z^\dagger(\pi)
&=J\hat S_x+\frac UN\hat S_z^2\\
&=-\hat H_{\rm BJJ}(-\Lambda).
\end{align}
Here $\hat R_z(\pi)$ denotes the collective rotations of $N$ indistinguishable spins about the z axis and $\hat H_{\rm BJJ}(\Lambda)$ denotes the Hamiltonian for a given $\Lambda$. In Eq.~(\ref{BJJ_rot_z}), we used the relationship $\hat R_z(\pi)\hat S_x\hat R_z^\dagger(\pi)=-\hat S_x$~\cite{Pezze}.

From Eq.~(\ref{BJJ_rot_z}), we have
\begin{align}\label{H(-Lambda)_1}
\hat H_{\rm BJJ}(-\Lambda)\hat R_z(\pi)|E_i(\Lambda)\rangle&=-E_i(\Lambda)R_z(\pi)|E_i(\Lambda)\rangle.
\end{align}
where $E_i(\Lambda)$ ($|E_i(\Lambda)\rangle$) is the energy eigenvalue (eigenstate) of the Hamiltonian $\hat H_{\rm BJJ}(\Lambda)$ for a given $\Lambda$.
On the other hand,
\begin{align}\label{H(-Lambda)_2}
\hat H_{\rm BJJ}(-\Lambda)|E_i(-\Lambda)\rangle=E_{i}(-\Lambda)|E_i(-\Lambda)\rangle.
\end{align}
Comparing Eq.~(\ref{H(-Lambda)_1}) with Eq.~(\ref{H(-Lambda)_2}), we have
\begin{gather}
\label{ev_Lambda}E_i(-\Lambda)=-E_{i}(\Lambda),\\
\label{es_Lambda}|E_i(-\Lambda)\rangle=R_z(\pi)|E_i(\Lambda)\rangle.
\end{gather}
Now let us expand $|E_i(\Lambda)\rangle$ as
\begin{align}\label{expand_E_i}
|E_i(\Lambda)\rangle=\sum_k^Nc_{i,k}(\Lambda)|N-k\rangle_{\rm L}|k\rangle_{\rm R},
\end{align}
where $c_{i,k}(\Lambda)=_{\rm L}\langle N-k|_{\rm R}\langle k|E_i(\Lambda)\rangle$. From Eqs.~(\ref{es_Lambda}) and (\ref{expand_E_i}), we have
\begin{align}\label{E}
|E_i(-\Lambda)\rangle=\sum_k^Nc_{i,k}(\Lambda)e^{-i(N-2k)\pi/2}|N-k\rangle_{\rm L}|k\rangle_{\rm R}.
\end{align}

When the initial state is set to be $|N-n\rangle_{\rm L}|n\rangle_{\rm R}$, the wave function for a given $\Lambda$ at time $t$ is written as
\begin{align}
|\Psi(\Lambda,t)\rangle=\sum_i^Nf_{n,i}(\Lambda)e^{-iE_i(\Lambda)t}\times\sum_k^Nc_{i,k}(\Lambda)|N-k\rangle_{\rm L}|k\rangle_{\rm R},
\end{align}
where $f_{n,i}(\Lambda)=\langle E_i(\Lambda)|N-n\rangle_{\rm L}|n\rangle_{\rm R}$. From Eq.~(\ref{es_Lambda}), we have the following mapping relationship between $f_{n,i}(\Lambda)$ and $f_{n,i}(-\Lambda)$:
\begin{align}\label{f}
f_{n,i}(-\Lambda)=e^{i(N-2n)\pi/2}f_{n,i}(\Lambda).
\end{align}
From Eqs. (\ref{E}) and (\ref{f}), we obtain
\begin{align}
|\Psi(-\Lambda,t)\rangle=\sum_i^N\sum_k^N(-1)^{k-n}f_{n,i}(\Lambda)e^{iE_{i}(\Lambda)t}\times c_{i,k}(\Lambda)|N-k\rangle_{\rm L}|k\rangle_{\rm R}.
\end{align}

Therefore, the probability of obtaining the result $|N-m\rangle_{\rm L}|m\rangle_{\rm R}$ at time $t$ for a given $\Lambda$, in the case of the initial state set to be $|N-n\rangle_{\rm L}|n\rangle_{\rm R}$, is then given by
\begin{align}\label{P_Lambda}
P_{m,n}(\Lambda,t)=\left|\sum_i^Nf_{n,i}(\Lambda)e^{-iE_i(\Lambda)t}c_{i,m}(\Lambda)\right|^2.
\end{align}
In contrast,
\begin{align}\label{P_-Lambda}
P_{m,n}(-\Lambda,t)=\qty|\sum_i^Nf_{n,i}(\Lambda)e^{iE_{i}(\Lambda)t}c_{i,m}(\Lambda)|^2.
\end{align}
From Eqs.~(\ref{P_Lambda}) and (\ref{P_-Lambda}), we have
\begin{align}
P_{m,n}(\Lambda,t)=P_{m,n}(-\Lambda,t).
\end{align}
Therefore, for any initial state $|N-n\rangle_{\rm L}|n\rangle_{\rm R}$, the probability of obtaining the result $|N-m\rangle_{\rm L}|m\rangle_{\rm R}$ at time $t$ is found to be symmetric between repulsive and attractive interaction.

\begin{figure}[b!]
\begin{subfigure}[t]{0.3\textwidth}
\includegraphics[width=\textwidth]{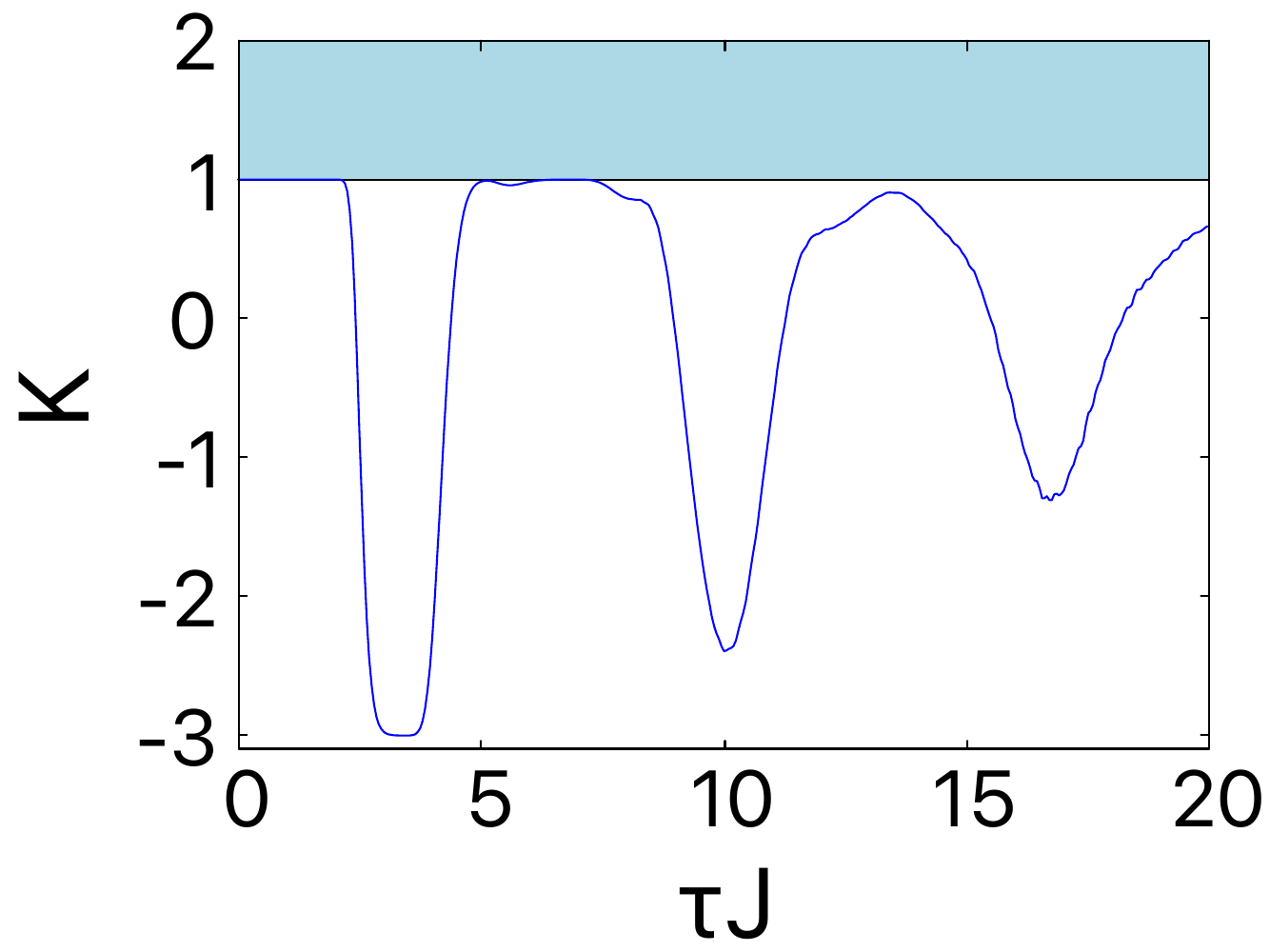}
\end{subfigure}
\caption{Time-evolution of $K$ under macrorealism for $N=100$ and $\Lambda=1$.}
\label{fig:LG_NIM} 
\end{figure}

\section{Calculation of the LGI for dynamics of bosons under macrorealism}\label{appendix:D}

We here consider $K$ for dynamics of bosons in a double-well potential under macrorealism. 

MRPS here assumes that the system possesses definite observables, which makes it possible to obtain $P(n_{\rm L})$ without collapsing the wave function. NIM here assumes that measurements at $t_2$ don't collapse the wave function, and as the result don't affect the subsequent dynamics. Under these two assumptions, calculating $P(n_{\rm L})$ using the quantum theory, $\ev{Q_2Q_3}$ of $K$ is given by
\begin{gather}\label{Q2Q3_NIM} 
    \ev{Q_2Q_3}=\sum_{n_{\rm L,2}}{\rm sgn}(n_{\rm L,2}-n_{\rm R,2})P(n_{\rm L,2},\tau)\sum_{n_{\rm L,3}}{\rm sgn}(n_{\rm L,3}-n_{\rm R,3})P(n_{\rm L,3},2\tau).
\end{gather}
It is important to emphasize that $P(n_{\rm L,3}, 2\tau)$ in Eq.~(\ref{Q2Q3_NIM}) is independent of $n_{\rm L,2}$, unlike $P(n_{\rm L,3},2\tau|n_{\rm L,2},\tau)$ in Eq.~(\ref{Q2Q3}). $\ev{Q_1Q_2}$ and $\ev{Q_1Q_3}$ of $K$ are given by Eqs.~(\ref{Q1Q2}) and (\ref{Q1Q3}), respectively. Since the initial state is set at $t_1$ and the dynamics after the measurement at $t_2$ ($t_3$) is not concerned when calculating $\ev{Q_1Q_2}$ ($\ev{Q_1Q_3}$), making the assumption of macrorealism does not change the result of $\ev{Q_1Q_2}$ ($\ev{Q_1Q_3}$). 

In this case, the LGI is satisfied, as shown in Fig.~\ref{fig:LG_NIM}. For the comparison with the the result of Fig. 2(a) in the main text, we here calculate $K$ for $N=100$ and $\Lambda=1$.

\section{Effect of measurement on dynamics}\label{appendix:E}
\begin{figure}[t!]
\begin{subfigure}[t]{0.3\textwidth}
\includegraphics[width=\textwidth]{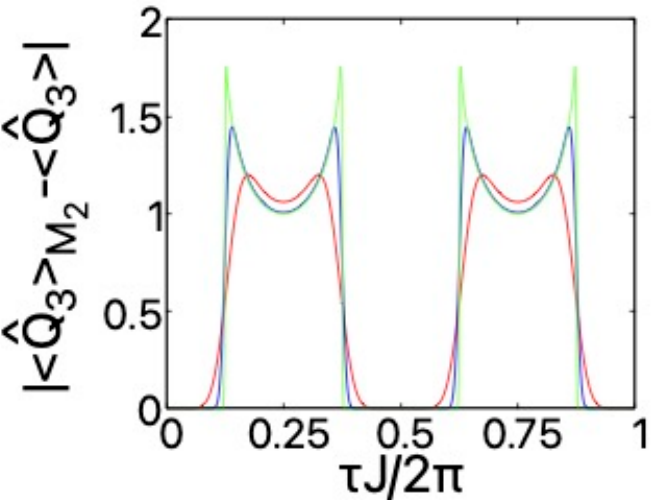}
\end{subfigure}
\caption{Difference between the mean value of $\hat Q_3$ with and without the measurement during the time evolution as a function of the time. 
The red line, blue line, and green line, respectively, correspond to the cases of $N=10$, $N=100$, and $N=10000$ with $U=0$. }
\label{fig:measurement} 
\end{figure}

As we have discussed in the main letter, the maximum value of $LG$ becomes larger with the increase of $N$. This fact is attributed to the effect of the measurement on $Q_3$.

In Fig.~\ref{fig:measurement}, we plot the difference of the mean value of $\hat Q_3$ for different $N$.
The red line, blue line, and green line, respectively, correspond to the cases of $N=10$, $N=100$, and $N=10000$ with $U=0$.
This quantity would characterize the effect of the measurement on the time evolution.
The figure shows that the effect of the measurement becomes more significant as $N$ becomes larger.

\section{Violation of the LGI for large particle number in the single particle regime}\label{appendix:F}

\begin{figure}[t!]
\centering
\begin{subfigure}[t]{0.30\textwidth}
\includegraphics[width=\textwidth]{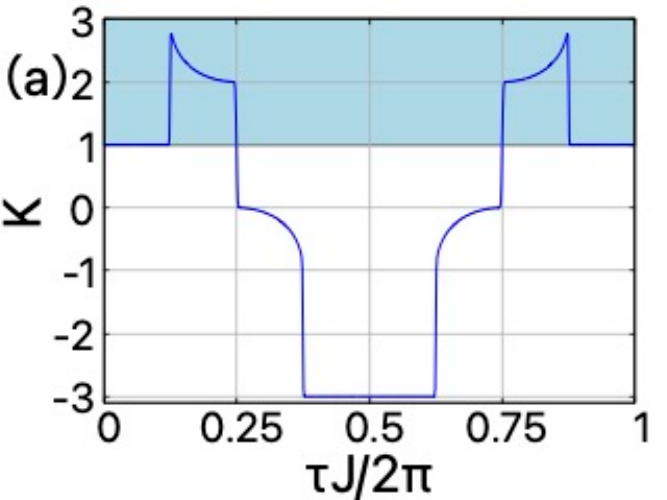}
\phantomsubcaption
\label{fig:LG_largeN} 
\end{subfigure}
\begin{subfigure}[t]{0.30\textwidth}
\includegraphics[width=\textwidth]{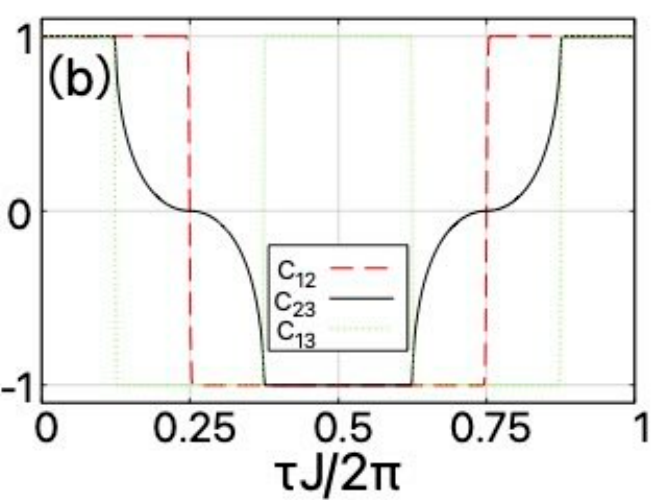}
\phantomsubcaption
\label{fig:TTC_largeN} 
\end{subfigure}
\caption{(a) Violation of the LGI. (b) Three correlation functions $C_{12}$ (the red dashed line), $C_{23}$ (the black solid line), and $C_{13}$ (the green dotted line) as functions of $\tau$. Here the parameters are $N=10000$ and $\Lambda=0$.}
\end{figure}

Here, we test the LGI for $N$ larger than $100$ in the absence of on-site interaction. Fig.~\ref{fig:LG_largeN} shows the violation of the LG inequality for $N = 10000$. Comparing it with the violation for $N = 100$ in Fig.~4(a) of the main text, it is clear that the value of the peak of $K$ becomes larger with the increase of $N$, due to more abrupt decrease of $C_{13}$ (see Fig.~\ref{fig:TTC_largeN}).

\section{Rabi oscillation regime}\label{appendix:G}

\begin{figure}[t!]
\centering
\begin{subfigure}[t]{0.3\textwidth}
\includegraphics[width=\textwidth]{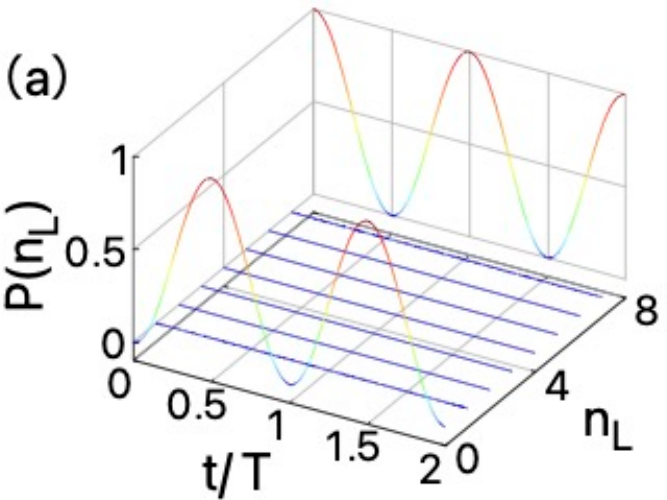}
\phantomsubcaption
\label{fig:rabi} 
\end{subfigure}
\begin{subfigure}[t]{0.3\textwidth}
\includegraphics[width=\textwidth]{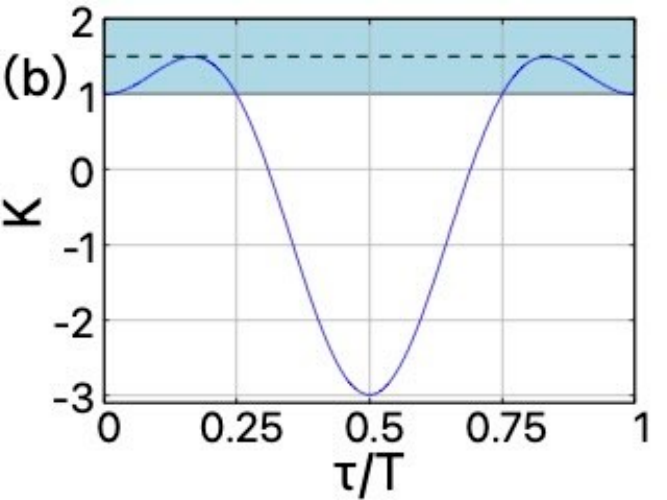}
\phantomsubcaption
\label{fig:rabi_vio} 
\end{subfigure}
\begin{subfigure}[t]{0.3\textwidth}
\includegraphics[width=\textwidth]{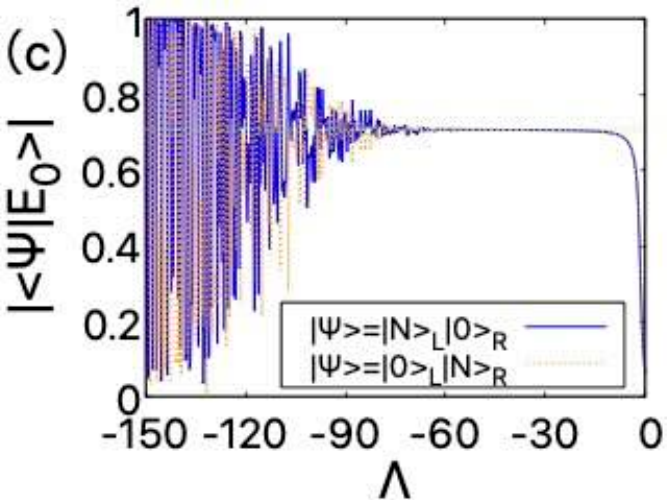}
\phantomsubcaption
\label{fig:gs_BJJ_N8} 
\end{subfigure}
\begin{subfigure}[t]{0.3\textwidth}
\includegraphics[width=\textwidth]{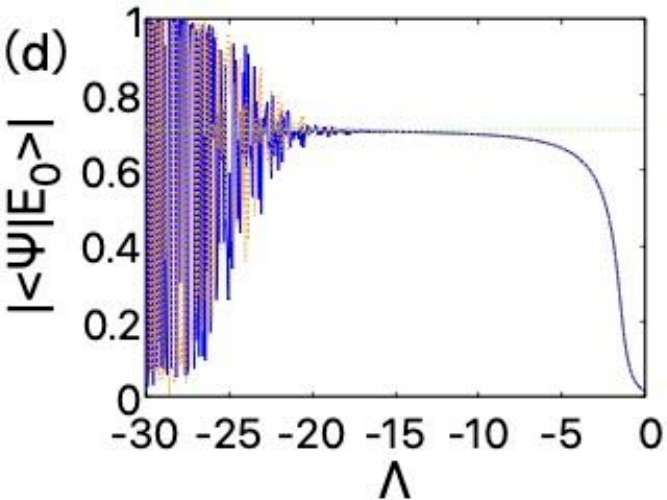}
\phantomsubcaption
\label{fig:gs_BJJ_N12} 
\end{subfigure}
\begin{subfigure}[t]{0.3\textwidth}
\includegraphics[width=\textwidth]{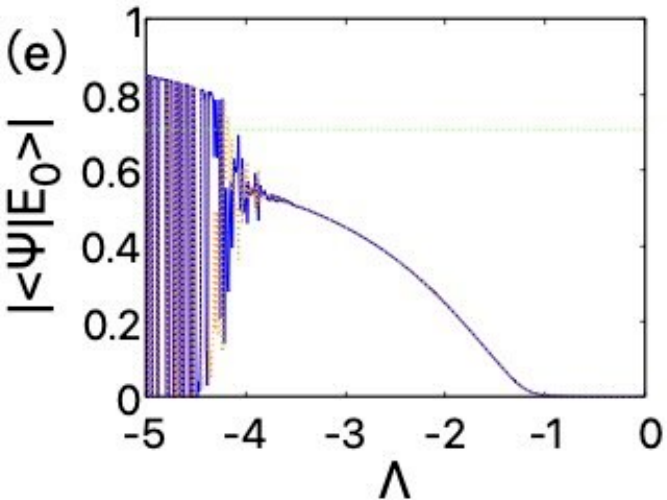}
\phantomsubcaption
\label{fig:gs_BJJ_N30} 
\end{subfigure}
\begin{subfigure}[t]{0.3\textwidth}
\includegraphics[width=\textwidth]{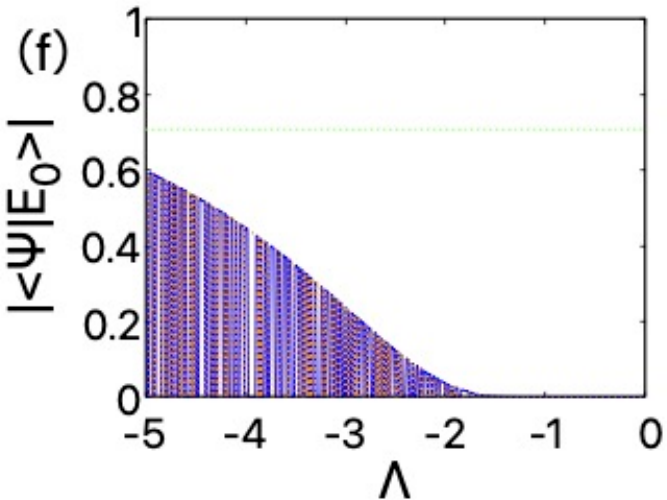}
\phantomsubcaption
\label{fig:gs_BJJ_N100} 
\end{subfigure}
\caption{(a) Two-mode oscillation between $|8\rangle_{\rm L}|0\rangle_{\rm R}$ and $|0\rangle_{\rm L}|8\rangle_{\rm R}$ for $\Lambda=-50$. 
(b) Violation of the LGI for the dynamics shown in (a). 
The dashed line indicates the maximum value ($K=1.5$).
$T$ denotes the period of the oscillation. 
(c)\textendash(f) Overlap between the ground state of $\hat H_{\rm BJJ}$ and $|N\rangle_{\rm L}|0\rangle_{\rm R}$ (the blue solid line) $|0\rangle_{\rm L}|N\rangle_{\rm R}$ (the orange dotted line) for a given $\Lambda$: $N=8$ (c), $12$ (d), $30$ (e), and $100$ (f). 
If the ground state is the bonding state, $\langle\Psi|E_0\rangle$ becomes $1/\sqrt2$ (the green dotted line).
}
\label{fig:gs_BJJ}
\end{figure}

We here consider the accessibility of the BJJ model to the Rabi-oscillation between $|N\rangle_{\rm L}|0\rangle_{\rm R}$ and $|0\rangle_{\rm L}|N\rangle_{\rm R}$ and show the violation of the LGI for the Rabi-oscillation.

It is known that there exists a regime where a mesoscopic two-mode oscillation takes place~\cite{Cirac, Dounas-Frazer}. 
Here we set the initial state to be $|N\rangle_{\rm L}|0\rangle_{\rm R}$, which gives $Q_1=1$. 
For a certain extent of the number of particles $N$, e.g.\,for $N=8$, the clear two-mode oscillation between $|N\rangle_{\rm L}|0\rangle_{\rm R}$ and $|0\rangle_{\rm L}|N\rangle_{\rm R}$ appears as shown in Fig.~\ref{fig:rabi}.

This oscillation is contributed only by the ground state $\ket{E_0}$ and the first excited state $\ket{E_1}$ of the Hamiltonian in the strong-coupling regime. 
Since, in this case, $\ket{E_0}\simeq(|N\rangle_{\rm L}|0\rangle_{\rm R}+|0\rangle_{\rm L}|N\rangle_{\rm R})/\sqrt2$ and $\ket{E_1}\simeq(|N\rangle_{\rm L}|0\rangle_{\rm R}-|0\rangle_{\rm L}|N\rangle_{\rm R})/\sqrt2$, the initial state can be approximated as a superposition of $\ket{E_0}$ and $\ket{E_1}$.
The dynamics is then given by
\begin{align}\label{BJJ_rabi}\tag{7}
|\Psi(t)\rangle\simeq\cos\omega_1t|N\rangle_{\rm L}|0\rangle_{\rm R}+i\sin\omega_1t|0\rangle_{\rm L}|N\rangle_{\rm R},
\end{align}
where, $\omega_1=(E_1-E_0)/2$. 
Here $E_0$ and $E_1$ are the eigenenergies for $\ket{E_0}$ and $\ket{E_1}$, respectively.
The period of the oscillation is then simply given by $T=\pi/\omega_1$. 
This oscillation breaks the LGI with a maximum value 1.5, as shown in Fig.~\ref{fig:rabi_vio}.  The Rabi oscillation, in fact, takes place without regard to the sign of $\Lambda$. When on-site interaction is repulsive, the degenerate highest energy eigenstates $\ket{E_8}$ and $\ket{E_7}$ contribute to Rabi oscillation in the same way as $\ket{E_0}$ and $\ket{E_1}$ do for strong attractive interaction: $\ket{E_8}\simeq(|N\rangle_{\rm L}|0\rangle_{\rm R}+|0\rangle_{\rm L}|N\rangle_{\rm R})/\sqrt2$, $\ket{E_7}\simeq(|N\rangle_{\rm L}|0\rangle_{\rm R}-|0\rangle_{\rm L}|N\rangle_{\rm R})/\sqrt2$.

For sufficiently large $N$, however, the two-mode oscillation disappears because the ground state no longer becomes the state $(|N\rangle_{\rm L}|0\rangle_{\rm R}+|0\rangle_{\rm L}|N\rangle_{\rm R})/\sqrt2$ anymore. 
As shown in Figs.~\ref{fig:gs_BJJ_N8}\textendash\ref{fig:gs_BJJ_N100}, for a certain extent of $N$, such as $N=8$ and $12$, 
there is the region of $\Lambda$ for which the ground state is $(|N\rangle_{\rm L}|0\rangle_{\rm R}+|0\rangle_{\rm L}|N\rangle_{\rm R})/\sqrt2$. 
However, when $N$ becomes larger, the ground state $|E_0\rangle$ no longer becomes $(|N\rangle_{\rm L}|0\rangle_{\rm R}+|0\rangle_{\rm L}|N\rangle_{\rm R})/\sqrt2$. 
It is consistent with the fact that self-trapping takes place for large $N$ beyond the critical value $\Lambda_{\rm c}$.

